\documentclass[conference]{IEEEtran}
\ifCLASSINFOpdf
\else
\fi

\usepackage{microtype}
\usepackage{graphicx}
\usepackage{subfigure}
\usepackage{booktabs} 
\usepackage{amsmath} 
\usepackage{multirow}
\usepackage{array}
\usepackage{pifont}
\newcommand{\cmark}{\ding{51}} 
\newcommand{\xmark}{\ding{55}} 

\begin{document}
%
\title{
    FlexLLM: Composable HLS Library for Flexible Hybrid LLM Accelerator Design
    \vspace{-0.3cm}
}



%
\author{
    \IEEEauthorblockN{
        Jiahao Zhang\IEEEauthorrefmark{1},
        Zifan He\IEEEauthorrefmark{1},
        Nicholas Fraser\IEEEauthorrefmark{2}, 
        Michaela Blott\IEEEauthorrefmark{2},
        Yizhou Sun\IEEEauthorrefmark{1}, 
        Jason Cong\IEEEauthorrefmark{1}
    }
    \IEEEauthorblockA{
        \IEEEauthorrefmark{1}Computer Science, University of California, Los Angeles, California\\ 
    }
    \IEEEauthorblockA{
        \IEEEauthorrefmark{2}AMD, Dublin, Ireland
    }
    \vspace{-1cm}
}


\maketitle

\begin{abstract}

We present FlexLLM, a composable High-Level Synthesis (HLS) library for rapid development of domain-specific LLM accelerators. FlexLLM exposes key architectural degrees of freedom for stage-customized inference, enabling hybrid designs that tailor temporal reuse and spatial dataflow differently for prefill and decode, and provides a comprehensive quantization suite to support accurate low-bit deployment. Using FlexLLM, we build a complete inference system for the Llama-3.2 1B model in under two months with only 1K lines of code. The system includes: (1) a stage-customized accelerator with hardware-efficient quantization (12.68 WikiText-2 PPL) surpassing SpinQuant baseline, and (2) a Hierarchical Memory Transformer (HMT) plug-in for efficient long-context processing. On the AMD U280 FPGA at 16nm, the accelerator achieves 1.29$\times$ end-to-end speedup, 1.64$\times$ higher decode throughput, and 3.14$\times$ better energy efficiency than an NVIDIA A100 GPU (7nm) running BF16 inference; projected results on the V80 FPGA at 7nm reach 4.71$\times$, 6.55$\times$, and 4.13$\times$, respectively. In long-context scenarios, integrating the HMT plug-in reduces prefill latency by 23.23$\times$ and extends the context window by 64$\times$, delivering 1.10$\times$/4.86$\times$ lower end-to-end latency and 5.21$\times$/6.27$\times$ higher energy efficiency on the U280/V80 compared to the A100 baseline. FlexLLM thus bridges algorithmic innovation in LLM inference and high-performance accelerators with minimal manual effort.

\end{abstract}


%
\IEEEpeerreviewmaketitle

\section{Introduction}

\begin{figure*}[t]
    \centering
    \includegraphics[width=1.0\linewidth]{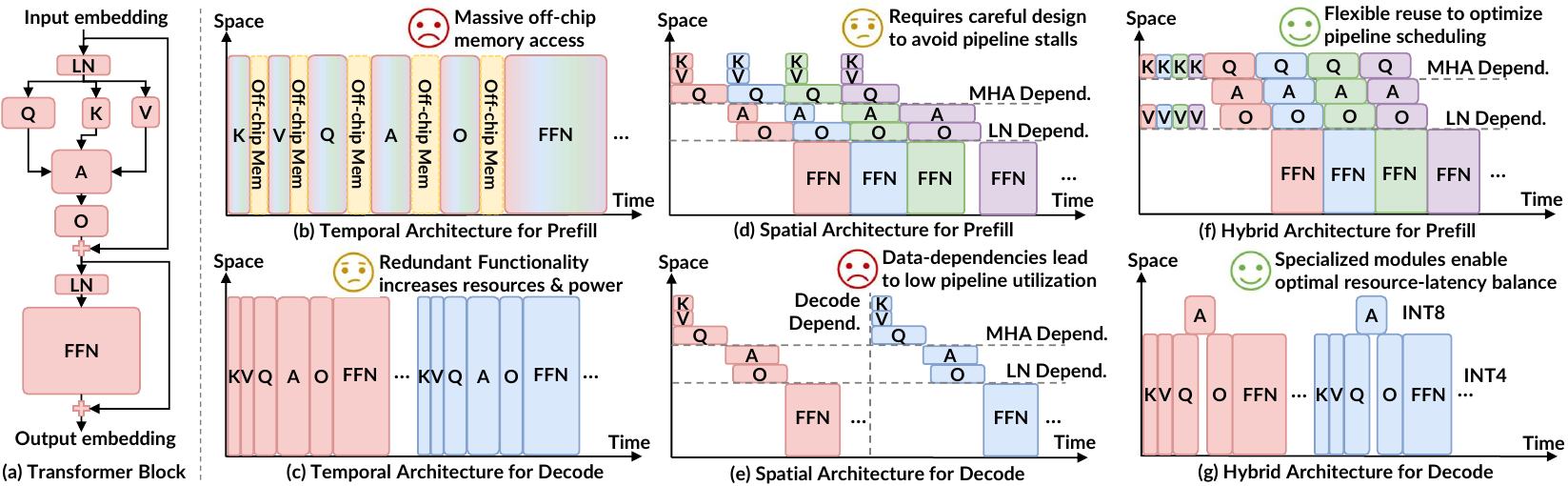}
    \vspace{-0.75cm}
    \caption{
        Comparison of three architectural styles for LLM accelerators across prefill and decode stages. Different colors indicate different tokens and blocks denote hardware modules. Only the linear layer is shown for clarity, where $A$ denotes the Multi-Head Attention (Grouped-Query Attention here) and $O$ denotes the output projection. 
        (a) One Transformer block. 
        (b–c) Temporal architectures achieve high utilization via module reuse but suffer from frequent off-chip memory access and limited flexibility. 
        (d–e) Spatial architectures dedicate modules per kernel for full on-chip streaming but are sensitive to pipeline stalls. 
        (f–g) Hybrid architectures combine temporal reuse and spatial parallelism to balance utilization, latency, and flexibility.
    }
    \label{fig:arch_comparison}
    \vspace{-0.3cm}
\end{figure*}

Large Language Models (LLMs) have emerged as one of the most transformative technologies of the modern era, driving breakthroughs across natural language processing~\cite{devlin2019bert, raffel2020t5}, code generation~\cite{chen2021codex, alphacode2022science}, and multimodal understanding~\cite{openai2023gpt4, alayrac2022flamingo}. While cloud providers continue investing heavily in GPU-based infrastructures to support large-scale inference~\cite{kwon2023vllm}, there is a growing demand for \textbf{domain-specific accelerators} that provide customized, efficient, and scalable alternatives across deployment environments—from data centers~\cite{hong2022dfx, chen2024understanding} to edge platforms~\cite{chen2019eyerissv2, huang2025edgellm}.

Compared to general-purpose CPUs and GPUs, domain-specific accelerators have demonstrated significant gains in performance and energy efficiency for prior DNN workloads (e.g., CNNs~\cite{zhang2016caffeine, basalama2023flexcnn}, ResNets~\cite{li2017fpga, ma2017end, guan2017fp}). However, modern LLMs differ substantially from conventional DNNs: they contain billions of parameters and exhibit more complex computation and memory access patterns~\cite{brown2020gpt3, dubey2024llama, liu2024deepseek}, resulting in far higher demands on compute throughput, memory bandwidth, and architectural flexibility. These characteristics present three key challenges in designing efficient LLM accelerators:

\textbf{Challenge 1: Divergent compute and memory behaviors in prefill vs.\ decode.}  
LLM inference with Transformer decoders consists of two stages with fundamentally different bottlenecks. During \emph{prefill}, the model processes the prompt in parallel, offering high arithmetic intensity and abundant parallelism. During \emph{decode}, tokens are generated autoregressively; computation is dominated by data dependencies and frequent memory accesses, making it strongly memory-bandwidth-bound. As a result, the same model is often compute-bound in prefill but memory-/dependency-bound in decode, creating conflicting optimization goals within one serving pipeline.

Existing FPGA accelerators primarily follow either \textbf{temporal architectures}~\cite{hong2022dfx, zeng2024flightllm} or \textbf{spatial architectures}~\cite{blott2018finn, fastml_hls4ml, chen2024allo, ye2025streamtensor}. Temporal designs reuse shared compute engines across layers, but incur frequent off-chip traffic in prefill due to limited buffering and struggle to support heterogeneous kernels/precisions within a single engine (Fig.~\ref{fig:arch_comparison}(b)(c)). Spatial designs map kernels to dedicated modules and stream intermediate data through on-chip FIFOs, but suffer from pipeline stalls when kernel latencies are unbalanced or when intrinsic dependencies dominate (Fig.~\ref{fig:arch_comparison}(d)(e)). Despite their differences, both paradigms typically share the same implicit choice: using a \emph{single unified architecture} to serve both stages.

However, since prefill and decode are bottlenecked by different resources, a unified (temporal, spatial, or even hybrid) design inevitably over-optimizes one stage while under-serving the other. In other words, \textbf{\textit{any ``one-size-fits-all'' architecture is fundamentally mismatched to LLM serving: prefill and decode must be stage-customized}}, with each stage adopting a different mix of spatial parallelism and temporal reuse to match its distinct compute/memory constraints.

\textbf{Challenge 2: Balancing model accuracy and low-bit compute/memory efficiency.}  
LLMs have enormous parameter counts, making compression indispensable for accelerator deployment~\cite{chen2024understanding, zeng2024flightllm}. Quantization is an effective way to reduce compute and memory cost; however, prior FPGA LLM accelerators often rely on naive integer quantization, which can severely degrade accuracy when pushed to the aggressive low-bit regime required to approach GPU-level throughput, making such designs impractical for real deployment. For example, applying INT4 SmoothQuant~\cite{xiao2023smoothquant} or GPTQ~\cite{frantar2022gptq} to Llama~3.2-1B increases perplexity to over $1e2$ on WikiText-2~\cite{liu2024spinquant}. Given FPGAs' limited compute capability relative to GPUs, \textbf{\textit{practical LLM acceleration must rely on state-of-the-art quantization with hardware-oriented co-design}}, rather than naive quantization pipelines.

\textbf{Challenge 3: High manual effort and slow iteration for model-specific accelerators.}  
As LLM architectures evolve rapidly, developing model-specific accelerators remains highly labor-intensive. RTL-based implementations can exceed 100K lines of code and impose a steep learning curve, making it difficult to iterate at the pace of LLM innovation. Recent HLS frameworks such as Allo~\cite{chen2024allo} and StreamTensor~\cite{ye2025streamtensor} reduce RTL burden via MLIR-based mapping and automated compilation, but they largely target spatial architectures and offer limited support for hybrid designs that optimize prefill and decode specifically.

To close this productivity gap, \textbf{\textit{a high-level programming framework that enables flexible architecture design is necessary for the rapid growth of LLM-oriented accelerators.}} Without a composable framework, accelerator development cycles (often 1--2 years) cannot keep pace with the rapid evolution of LLMs, where new model releases occur on the order of months. What is needed is a framework that exposes the key architectural degrees of freedom—especially stage-customized hybrid design and advanced quantization—while remaining accessible to both ML and accelerator developers.

Motivated by these challenges, we develop \textbf{FlexLLM}, a composable HLS library for domain-specific LLM accelerator design. Built on TAPA~\cite{guo2023tapa}, FlexLLM provides highly parameterized and templated modules that enable rapid construction of hybrid accelerators. Importantly, FlexLLM is, to our knowledge, \textbf{\textit{the first to explicitly go beyond the unified-design paradigm and enable stage-customized hybrid architectures}}. This significantly improves design flexibility and accelerator performance while offering a reusable methodology for future LLM accelerator development. Using FlexLLM, we build a high-performance accelerator system that integrates state-of-the-art LLM techniques with fewer than 1K lines of C++ code, demonstrating that FlexLLM can rapidly incorporate new algorithmic innovations while significantly reducing development effort. We will open-source FlexLLM upon publication.   

In summary, our main contributions include:
\begin{itemize}
    \item We present FlexLLM, a composable HLS library that supports \textit{flexible hybrid, stage-customized architecture construction} for LLM accelerators. Its highly templated modules enable rapid model-specific customization while significantly reducing design complexity and code volume.
    \item We integrate a comprehensive quantization stack---covering dynamic and static variants, multiple symmetry and granularity options, and outlier-handling modules---enabling accurate and efficient LLM deployment on customized accelerator platforms. To the best of our knowledge, this provides the most advanced quantization support among existing LLM accelerator frameworks.
    \item Leveraging FlexLLM, we build a stage-customized hybrid accelerator for Llama-3.2~1B with a hardware-aware W4A4KV8 SpinQuant~\cite{liu2024spinquant} scheme, reducing WikiText-2 PPL from 13.30 (original SpinQuant) to 12.68. Compared to the BF16 baseline on an NVIDIA A100 GPU, our FPGA implementation delivers 1.29$\times$ end-to-end speedup, 1.64$\times$ higher decode throughput, and 3.14$\times$ better energy efficiency on AMD Alveo U280, and an estimated 4.71$\times$, 6.55$\times$, and 4.13$\times$ gains, respectively, on Versal V80.
    \item We further implement a Hierarchical Memory Transformer (HMT)~\cite{he2024hmt} plug-in on top of FlexLLM for long-context processing, reducing prefill latency by up to 23.23$\times$ and extending the effective context window by over 64$\times$ with less than 7.5\% resource and 0.6\% latency overhead, demonstrating FlexLLM’s scalability to long-context acceleration.
\end{itemize}

\section{Background}

\subsection{Characterization of LLM Inference Stages}
\label{ssec:llm-inference-stages}

The Transformer architecture~\cite{vaswani2017attention} forms the backbone of most modern LLMs. Each Transformer layer primarily consists of a multi-head self-attention (MHA) block and a feed-forward network (FFN), followed by normalization and residual connections, as shown in Fig.~\ref{fig:arch_comparison}(a). This highly regular structure makes a kernel-based hardware library a natural and efficient choice for accelerator design.

During inference, decoder-only LLMs are typically executed in two stages: prefill and decode. In prefill, the model processes the entire prompt in parallel and computes activations for all tokens, exhibiting high arithmetic intensity and abundant inter-token parallelism, and is thus largely compute-bound. In decode, the model generates tokens autoregressively while reusing the Key--Value (KV) cache; computation becomes dominated by sequential dependencies and KV access, shifting the bottleneck from compute throughput to memory bandwidth.

\begin{figure}[t]
    \centering
    \includegraphics[width=0.75\linewidth]{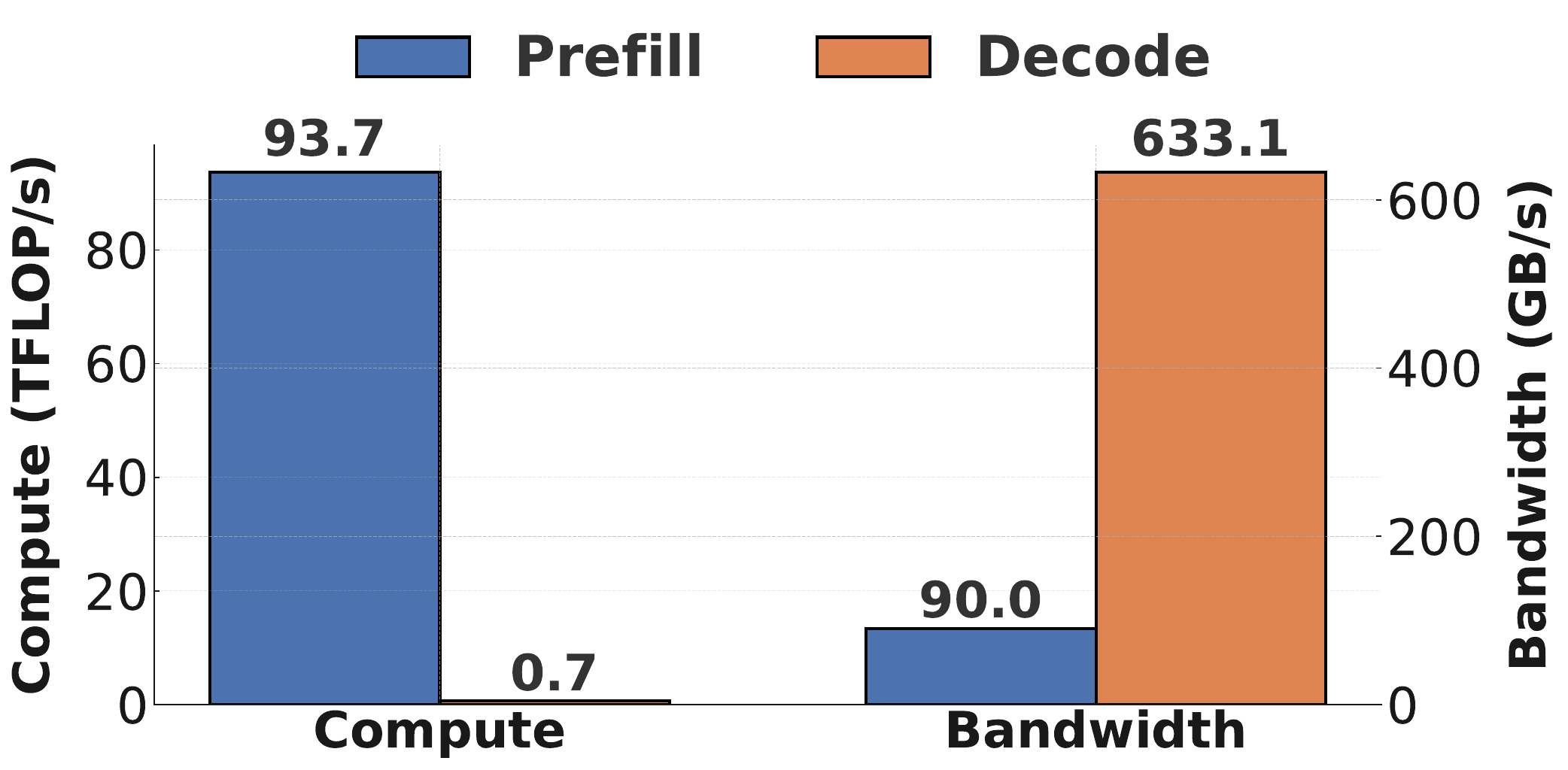}
    \vspace{-0.3cm}
    \caption{Profiling of compute throughput and memory bandwidth utilization during the prefill and decode stages of the BF16 Llama-3.2 1B model on an NVIDIA A100 GPU.}
    \label{fig:prefill_decode_compute_bandwidth}
    \vspace{-0.4cm}
\end{figure}

To quantify this stage divergence, we run BF16 Llama-3.2~1B~\cite{hf_llama_3_2_1b_access} on an NVIDIA A100 with vLLM~\cite{kwon2023vllm} and measure compute and memory-bandwidth utilization when processing and generating 1K tokens in prefill and decode, respectively (Fig.~\ref{fig:prefill_decode_compute_bandwidth}). The GPU achieves high utilization and strong performance in prefill due to massive parallel compute and highly optimized kernels. In contrast, during decode, compute utilization drops sharply and memory bandwidth becomes the dominant limiter, confirming that the two stages stress fundamentally different resources.

Table~\ref{tab:hardware_comparison} compares representative FPGA platforms~\cite{amd_u280_ds, amd_versal_v80_ds} with the A100 GPU~\cite{nvidia_a100_80gb_pcie_ds}. Notably, the decode stage on GPU often fails to fully utilize its peak bandwidth and compute, whereas modern FPGAs provide ample peak compute and HBM bandwidth to meet decode demands when memory traffic is reduced by quantization. Consequently, with effective low-bit quantization and an architecture co-designed for memory reuse and pipeline efficiency, domain-specific accelerators can be highly competitive—especially for decode-intensive, long-sequence generation workloads where token latency and energy efficiency are critical~\cite{chen2024understanding, joel2024survey, gomez2023confederacy}. 

\begin{table}[t]
\centering
\caption{Comparison of key hardware metrics across NVIDIA A100 GPU, AMD Alveo U280 FPGA, and Versal V80 FPGA.}
\resizebox{1.0\linewidth}{!}{%
    \begin{tabular}{l|c|c|c}
    \hline
    \textbf{Metric}     & \textbf{U280} & \textbf{V80}  & \textbf{A100 80GB PCle} \\
    \hline
    Technology Node     & TSMC 16nm         & TSMC 7nm              & TSMC 7nm   \\
    Peak Compute        & 8 FP32 TFLOPS     & 58 FP32 TFLOPS      & 312 FP32 TFLOPS   \\
    Peak HBM BW         & 460 GB/s          & 820 GB/s              & 1,935 GB/s    \\
    HBM Capacity        & 8 GB              & 32 GB                 & 80 GB         \\
    Peak Power          & 75 W             & 190 W                 & 300 W         \\
    \hline
    \end{tabular}%
}
\label{tab:hardware_comparison}
\vspace{-0.2cm}
\end{table}

\begin{table*}[t]
\centering
\caption{Comparison of related accelerators and frameworks.}
\label{tab:framework-comparison}
\vspace{-0.2cm}
\resizebox{\textwidth}{!}{
    \begin{tabular}{lcccccc}
    \hline
    Accelerator Framework &
    Contribution Type &
    Arch. Type &
    Extensibility &
    stage-customized kernel Opt. &
    Support Precision &
    Support Quant. Type \\
    \hline
    FlightLLM~\cite{zeng2024flightllm} & Accelerator & Temporal & Not extensible & \xmark & FP32, INT1$\sim$8 & Static \\
    FINN~\cite{blott2018finn}, hls4ml~\cite{fastml_hls4ml} & Compiler Framework & Spatial & Automatic Mapping & \xmark & FP32, INT1$\sim$8 & Static \\
    Allo~\cite{chen2024allo} & Domain-Specific Language & Spatial & Require Manual Design & \xmark & FP32, INT4/8 & Static \\
    StreamTensor~\cite{ye2025streamtensor} & Compiler Framework & Spatial & Automatic Optimization & \xmark & FP32, INT8 & Static \\
    SSR~\cite{zhuang2024ssr}, InTAR~\cite{he2025intar} & Accelerator & Hybrid & Require Manual Design & \xmark & FP32, INT8 & Static \\
    
    \hline
    FlexLLM (Ours) & HLS Kernel Library & Hybrid & Templated Kernel Composition & \cmark & FP16/32, INT1$\sim$8 & Static/Dynamic \\
    \hline
    \end{tabular}%
}
\vspace{-0.4cm}
\end{table*}

\subsection{Hardware-efficient LLM Quantization}

Quantization is essential for efficient LLM deployment on domain-specific accelerators. A general $N$-bit integer quantization can be written as
\[
X_q = s \cdot \text{round}\left(\frac{X - b}{s}\right) + b,
\]
where $X$ is the floating-point (FP) tensor, $X_q$ is the quantized integer tensor, and $s$ and $b$ are the scale and zero offset. Symmetric quantization uses $s = \frac{\max(|X|)}{2^{N-1}-1}$ and $b = 0$, while asymmetric quantization uses $s = \frac{\max(X) - \min(X)}{2^{N}-1}$ and $b = \min(X)$. Quantization is \emph{static} when $s,b$ are precomputed offline, and \emph{dynamic} when they are measured at runtime.

Static quantization is naturally compatible with pre-trained weights and is hardware-friendly for activations since it avoids runtime calibration, and thus has been widely adopted in recent accelerators~\cite{zeng2024flightllm,chen2024understanding}. However, its accuracy often degrades sharply at INT4, even with advanced schemes such as SmoothQuant~\cite{xiao2023smoothquant}. In contrast, fine-grained dynamic quantization introduces extra overhead and token-wise dependency barriers, but becomes a practical route to accurate aggressive low-bit deployment when combined with outlier-handling techniques, such as activation rotation and Fast Hadamard Transform (FHT) introduced in SpinQuant~\cite{liu2024spinquant}.

\subsection{Related Accelerators and Frameworks}

Table~\ref{tab:framework-comparison} summarizes representative Transformer-oriented accelerators and frameworks.

FlightLLM~\cite{zeng2024flightllm} adopt a \textbf{temporal} design that sequentially executes layers on shared engines to maximize reuse. While utilization can be high, limited on-chip buffering incurs frequent off-chip accesses, and the monolithic engine limits support for heterogeneous kernels/precisions and rapid model customization, increasing latency, energy, and redesign effort. 

In contrast, FINN~\cite{blott2018finn}, hls4ml~\cite{fastml_hls4ml}, Allo~\cite{chen2024allo} and StreamTensor~\cite{ye2025streamtensor} embrace a fully \textbf{spatial-dataflow} paradigm, mapping modules to dedicated hardware and streaming data on-chip. They provide useful abstractions or compilation support for customization, but decode performance remains limited by pipeline stalls under intrinsic dependencies and unbalanced kernel latencies, where throughput becomes hard to sustain.

Hybrid architectures aim to combine the strengths of temporal and spatial designs via flexible scheduling and module-level customization. For example, SSR~\cite{zhuang2024ssr} and InTAR~\cite{he2025intar} instantiate a fixed set of linear and non-linear engines, enabling inter-module streaming and intra-engine kernel switching. However, existing hybrids largely retain a \emph{unified} architecture across inference stages; fixed engine configurations restrict deeper optimization across diverse LLM models and the sharply different bottlenecks of prefill and decode.

To fully realize the potential of hybrid LLM acceleration, \textbf{stage-customized} design is necessary, where prefill and decode adopt different module configurations and scheduling policies to match their distinct hardware characteristics (Fig.~\ref{fig:arch_comparison}(f)(g)). In the next section, we describe how FlexLLM achieve this goal through stage-oriented module optimization and rapid hybrid construction, while integrating broad quantization support to build practical high-performance LLM accelerators with a model-to-silicon turnaround of less than two months.

\section{FlexLLM Framework}
\subsection{stage-customized Module Templates}

As discussed in ~\ref{ssec:llm-inference-stages}, prefill and decode stress fundamentally different resources, so a single unified architecture inevitably compromises one stage. FlexLLM addresses this by providing \textbf{stage-customized module templates} that expose the right parallelism knobs for each stage, enabling rapid construction of hybrid accelerators with balanced pipelines.

\begin{figure}[t]
    \centering
    \includegraphics[width=0.9\linewidth]{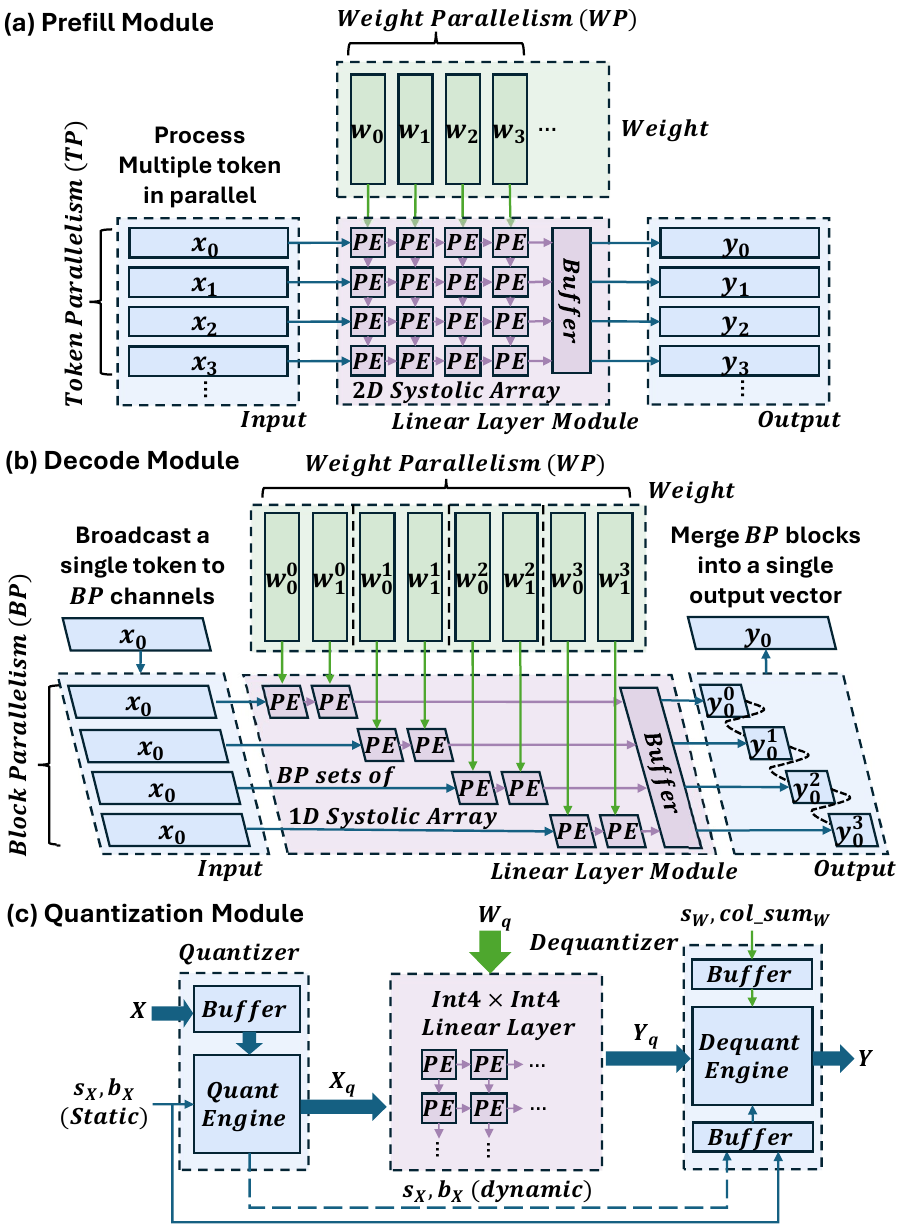}
    \vspace{-0.3cm}
    \caption{stage-customized module design for (a) prefill, (b) decode, and (c) quantization module integration.}
    \label{fig:moudle}
    \vspace{-0.4cm}
\end{figure}

\textbf{Prefill-stage modules:}  
Prefill exposes abundant inter-token parallelism. FlexLLM packs activations from multiple tokens and processes them concurrently for both linear and non-linear layers (Fig.~\ref{fig:moudle}(a)). We define this inter-token parallelism as \textit{\textbf{token\_parallelism (TP)}}. Since linear layers dominate compute, we further introduce \textit{\textbf{weight\_parallelism (WP)}} to fetch and stream multiple weight channels from off-chip memory in parallel. Each prefill linear module implements a 2D systolic array of $\textit{TP}\times\textit{WP}$ processing elements (PEs), optimized with an initiation interval (II) of one cycle for supported precisions. 

Given a prompt of length $l_{p}$, the theoretical prefill latency of a linear layer with input dimension $d_{in}$ and output dimension $d_{out}$ can be expressed as
\begin{equation}
T^{p}_{\text{linear}}=\frac{l_p\, d_{in}\, d_{out}}{\textit{TP}\cdot\textit{WP}},
\label{eq:linear_latency_prefill}
\end{equation}
and the corresponding off-chip bandwidth demand is
\begin{equation}
BW^{p}_{\text{linear}} = B_W \cdot \textit{WP} \cdot F,
\label{eq:linear_bandwidth}
\end{equation}
where $B_W$ is bytes per weight element and $F$ is the operating frequency. These knobs allow users to balance layer throughput while respecting memory-bandwidth constraints.

\textbf{Decode-stage modules: }  
Decode is constrained by autoregressive dependencies, leaving parallelism primarily within a single token. FlexLLM therefore exploits intra-token parallelism by partitioning each output hidden vector into blocks computed in parallel and reduced on-chip. We denote this as \textit{\textbf{block\_parallelism (BP)}}. Each decode linear module also supports \textit{\textbf{WP}}, implemented as $\textit{BP}$ sets of 1D systolic arrays with $\textit{WP}/\textit{BP}$ PEs each (Fig.~\ref{fig:moudle}(b)). For decode length $l_d$, the idea latency of a linear layer with the same $d_{in}$ and $d_{out}$ is
\begin{equation}
T^{d}_{\text{linear}}=\frac{l_d\, d_{in}\, d_{out}}{\textit{WP}}.
\label{eq:linear_latency_decode}
\end{equation}
The off-chip bandwidth follows Eq.~\ref{eq:linear_bandwidth}. Compared to prefill, decode PEs do not share weights across tokens, so the same resource budget can often support a higher $\textit{WP}$, increasing bandwidth demand and motivating careful stage-specific tuning.

\textbf{Quantization modules: }
Quantization modules use the same configurable parallelism (\textit{TP} for prefll and \textit{BP} for decode) as the non-linear modules. As shown in Figure~\ref{fig:moudle}(c), the quantizer converts FP inputs to low-bit integers using scales and zero offsets, which are either preloaded (static) or computed online (dynamic). After kernel computation, the dequantizer reconstructs FP outputs using the same scales and offsets, along with auxiliary data (e.g., per-channel weight scales and sums) buffered on-chip. In general, our framework supports static/dynamic and symmetric/asymmetric quantization with per-tensor, per-token, and per-channel granularities, and includes outlier-handling modules such as rotation and FHT.

\textbf{Library components: }
Following these principles, FlexLLM provides a comprehensive module library built on TAPA, comprising over 10K lines of highly parameterized code. It includes core LLM kernels, quantization/dequantization modules, and auxiliary components for on-chip streaming, buffering, and memory access management. Table~\ref{tab:module_template} summarizes the key modules and configurable parameters. These templates enable rapid, composable hybrid architecture exploration while remaining accessible to non-expert developers, reducing model-to-silicon turnaround from months to weeks.

\begin{table*}[t]
\centering
\scriptsize
\caption{Overview of primary module templates, configurable parameters, and Module interfaces provided in FlexLLM.}
\vspace{-0.2cm}
\renewcommand{\arraystretch}{1.1}
\setlength{\tabcolsep}{3.4pt}
\begin{tabular}{m{1cm}|m{1.8cm}|m{9.4cm}|m{4.8cm}}
\hline
 & \textbf{Module Template} & \textbf{Module Configurable Parameter} & \textbf{Module Interface} \\
\hline
\multirow{3}{=}{\raggedright\arraybackslash \textbf{Kernel Library}} 
& Linear Layer & 
\texttt{dtype, token\_parallelism(prefill), block\_parallel(decode), head\_parallel(MHA), weight\_parallelism, head\_num(MHA), max\_in\_dim, max\_out\_dim, max\_seq\_len} &
\texttt{in\_stream, w\_stream, out\_stream, in\_dim, out\_dim, seq\_len} \\
\cline{2-4}
& Non-Linear Layer (RoPE, Softmax, LayerNorm, ...) &
\texttt{dtype, token\_parallelism(prefill), block\_parallel(decode), head\_parallel(MHA), head\_num(MHA), max\_io\_dim, max\_seq\_len} &
\texttt{in\_stream, out\_stream, io\_dim, seq\_len} \\
\hline
\multirow{3}{=}{\raggedright\arraybackslash \textbf{Quant Library}}
& Static/Dynamic Quant Layer &
\texttt{in\_dtype, in\_quant\_bit, token\_parallelism(prefill), block\_parallel(decode), head\_parallel(MHA), in\_quant\_type(sym/asym), in\_quant\_granularity(per-tensor/token), head\_num(MHA), max\_io\_dim, max\_seq\_len} &
\texttt{in\_stream, in\_scale\_stream, in\_zero\_stream, quant\_in\_stream, io\_dim, seq\_len} \\
\cline{2-4}
& Static/Dynamic Dequant Layer &
\texttt{in\_quant\_bit, w\_quant\_bit, out\_dtype, token\_parallelism(prefill), block\_parallel(decode), head\_parallel(MHA), in\_quant\_type(sym/asym), in\_quant\_granularity(per-tensor/token), w\_quant\_type(sym/asym), w\_quant\_granularity(per-tensor/channel), head\_num(MHA), max\_io\_dim, max\_seq\_len} &
\texttt{quant\_out\_stream, in\_scale\_stream, in\_zero\_stream, w\_scale\_stream, w\_col\_sum\_stream, out\_stream, io\_dim, seq\_len} \\
\hline
\end{tabular}
\label{tab:module_template}
\vspace{-0.4cm}
\end{table*}

\subsection{Hybrid Accelerator Construction \& Optimization}

\begin{figure}[t]
    \centering
    \includegraphics[width=0.8\linewidth]{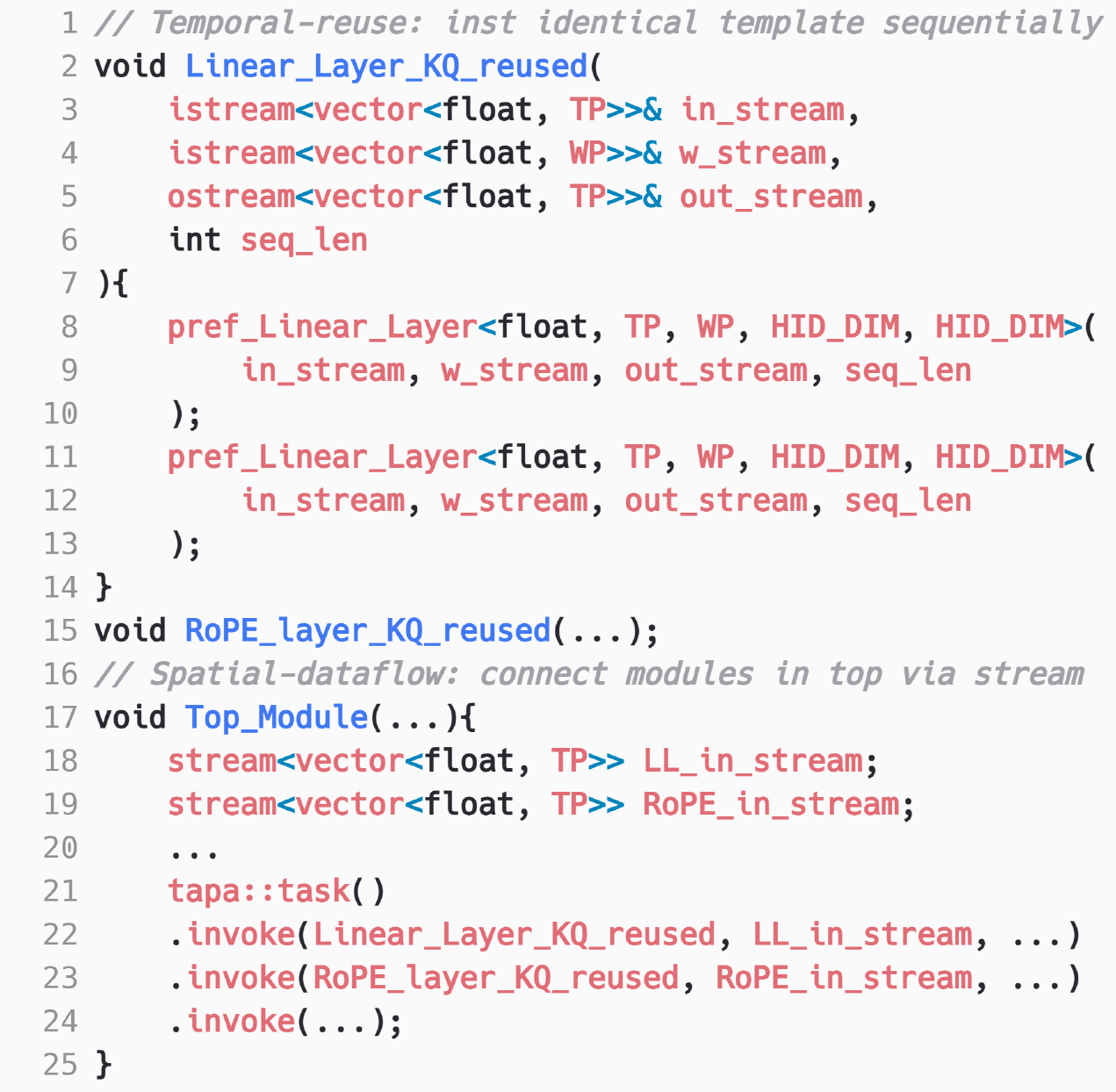}
    \vspace{-0.4cm}
    \caption{Example code illustrating hybrid architecture construction combining temporal reuse and spatial dataflow with FlexLLM.}
    \label{fig:code}
    \vspace{-0.8cm}
\end{figure}

Built upon the FlexLLM and TAPA HLS flows, users can efficiently explore hybrid architectures for LLM accelerators and quickly transition from design exploration to on-board implementation. Figure~\ref{fig:code} illustrates a simplified example that composes several modules in a hybrid style.

In the temporal-reuse part, the same templated module is instantiated and invoked multiple times within a single function to sequentially process similar operations. In this example, the prefill module of the linear layer and RoPE is reused for both \textit{Key} and \textit{Query} computations. This approach maximizes hardware utilization by reusing computation resources across similar tasks while minimizing redundant module instantiations.

In the spatial dataflow part, each instantiated module is explicitly invoked at the top level and connected through on-chip FIFOs to enable parallel execution and streaming communication. This module-based composition exploits inter-module pipelining and on-chip dataflow to achieve high throughput and overlap between computation stages.

The combination of temporal and spatial styles can effectively balance performance and resource efficiency in hybrid accelerator design. We illustrate a detailed methodology for stage-customized hybrid design through a case study in Sec.~\ref{case_study_1}. After composing an accelerator with FlexLLM, users can leverage its seamless compatibility with AutoBridge~\cite{guo2021autobridge} to optimize placement and routing (P\&R), enabling parallel exploration of design candidates to achieve higher frequency and performance. This end-to-end toolchain accelerates iteration and supports efficient on-board deployment.

\begin{figure*}[t]
    \centering
    \includegraphics[width=0.9\linewidth]{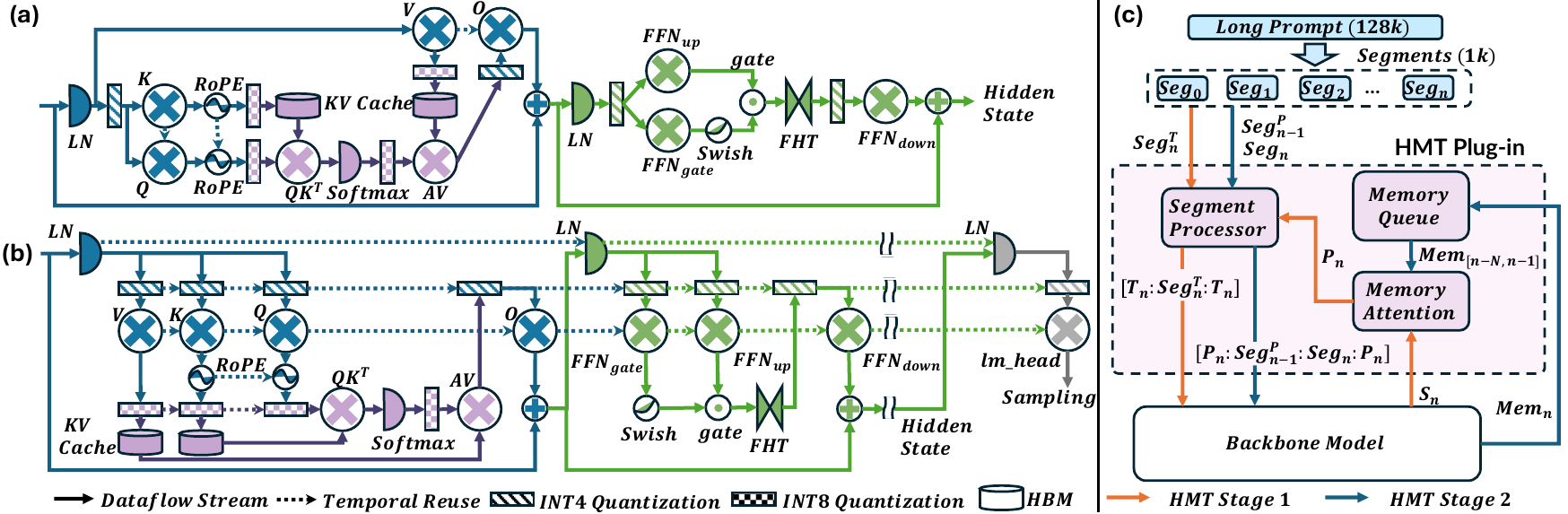}
    \vspace{-0.2cm}
    \caption{Hybrid architecture design for (a) prefill \& (b) decode, and  (c) the integration diagram of HMT plug-in.}
    \label{fig:Implementation}
    \vspace{-0.4cm}
\end{figure*}

\section{Case Study 1: Llama 3.2-1B + SpinQuant}
\label{case_study_1}


In this section, we demonstrate how FlexLLM is utilized to construct a high-performance accelerator with stage-customized architectures for the Llama-3.2 1B model. Specifically, we first refine the SpinQuant framework to obtain a more hardware-efficient representation while minimizing accuracy degradation. We then build customized architectures for prefill and decode with composable modules from FlexLLM. Table~\ref{tab:module_summary} summarizes the major module components, along with their corresponding lines of code and estimated development workload.

\subsection{Hardware-Efficient Model Quantization}

SpinQuant improves low-bit accuracy by mitigating activation outliers via learned rotations~\cite{liu2024spinquant}. While many rotations can be folded into weights or implemented efficiently with FHT, the remaining boundary rotations still require FP compute comparable to a full linear layer, which is costly on FPGAs. Moreover, the baseline SpinQuant setup keeps the MHA query path and the final vocabulary projection (\texttt{lm\_head}) in FP, leaving substantial hardware efficiency unrealized.

To obtain a practical model with minimal accuracy loss, we make the following hardware-oriented refinements:

\textbf{Remove boundary rotations.}  
We eliminate the first and last rotations in each decoder layer by absorbing them into adjacent weights during fine-tuning. This preserves the outlier-mitigation effect while removing FP boundary operations entirely.

\textbf{Quantize attention with higher precision.}  
Since MHA is more precision-sensitive, we quantize attention to INT8 while keeping other linear layers at INT4. For hardware simplicity, MHA uses static symmetric per-tensor quantization, while the remaining linear layers use dynamic asymmetric per-token quantization to retain accuracy. 

\textbf{Integer vocabulary projection.}  
We further quantize \texttt{lm\_head} to INT4, matching the other linear layers, reducing resource cost and improving decode throughput.

Table~\ref{tab:ablation} reports an ablation on WikiText-2. Starting from BF16 (PPL = 8.94), the original SpinQuant setup (INT4 linear layers with BF16-INT4 attention) yields PPL = 13.30. Raising attention precision to INT8 recovers most of the lost accuracy; using static INT8 slightly relaxes PPL but simplifies hardware. Finally, quantizing \texttt{lm\_head} increases PPL only marginally while enabling a fully integer pipeline across all linear layers.

\subsection{Architecture Design}

Using FlexLLM's stage-customized modules, we implement separate architectures for prefill and decode to match their distinct bottlenecks and achieve better performance and efficiency.

\textbf{Prefill stage:}  
Prefill benefits from inter-token parallelism and pipelined streaming. Unlike a purely spatial design that instantiates a dedicated module for each kernel, our prefill architecture combines streaming dataflow with selective temporal reuse. Shown as Fig.~\ref{fig:Implementation}(a), we share the same linear and RoPE modules between \texttt{Q\_proj} and \texttt{K\_proj}, and similarly for \texttt{V\_proj} and \texttt{O\_proj}. We compute \textit{Key} and \textit{Value} in parallel and store them in off-chip HBM, after which the remaining kernels execute in a fully pipelined dataflow manner. Streaming links maximize overlap across tokens, while temporal reuse mitigates idleness caused by data dependencies and helps form a balanced pipeline in MHA.

Following Eq.~\ref{eq:linear_latency_prefill}, the prefill latency bound is
\begin{equation} \label{eq:prefill_latency}
    T_{p} = 
\frac{Nl_{p}}{\textit{TP}} ( 
\frac{d_{h}d_{\textit{kv}}}{\textit{WP}_{\textit{kqvo}}} + 
\max( 
\frac{d_{h}^2}{\textit{WP}_{\textit{kqvo}}}, 
\frac{d_{h}l_{p}}{\textit{WP}_{\textit{mha}}}, 
\frac{d_{h}d_{\textit{ffn}}}{\textit{WP}_{\textit{ffn}}}))
\end{equation}
where $N$ is the number of decoder layers, and $d_h$, $d_{\textit{kv}}$, and $d_{\textit{ffn}}$ are the hidden, KV, and FFN intermediate dimensions. $\textit{WP}_{\textit{kqvo}}$, $\textit{WP}_{\textit{mha}}$, and $\textit{WP}_{\textit{ffn}}$ denote the weight-parallelism settings of the corresponding modules. The peak bandwidth demand is
\begin{equation}\label{eq:prefill_bandwidth}
BW_{p}=F\!\left(B_W^{\textit{int4}}\!\left(2\textit{WP}_{\textit{kqvo}}+3\textit{WP}_{\textit{ffn}}\right)+B_W^{\textit{int8}}\cdot 2\textit{WP}_{\textit{mha}}\right)
\end{equation}
where $B_W$ is bytes per weight element and F denotes frequency.

In prefill, non-linear overheads scale mainly with \textit{TP}, while linear modules scale with both \textit{TP} and \textit{WP}. We tune these parameters via Integer Linear Programming (ILP) under hardware constraints (resources and memory bandwidth) to minimize $T_{p}$ and obtain an optimal, well-pipelined design.

\textbf{Decode stage:}  Autoregressive dependencies prevent inter-token parallelism in decode, making spatial overlap across tokens infeasible. To maximize runtime hardware utilization, we temporally reuse modules for all INT4 quantization, linear layers, and non-linear layers, while maintaining dataflow connectivity within MHA and between reused quantization, linear, non-linear modules, shown as Fig.~\ref{fig:Implementation}(b). This design enables intra-token parallelism and inter-head overlay, effectively exploiting available concurrency.

The theoretical latency bound and memory bandwidth of the decode stage are given in Eq.~\ref{eq:decode_latency} and Eq.~\ref{eq:decode_bandwidth}, where $\textit{WP}_{int4}$ denotes the shared \textit{WP} used for the projection, FFN, and \texttt{lm\_head} layers. We similarly tune \textit{BP}s and \textit{WP}s via ILP under hardware constraints to minimize $T{d}$ on the target device.
\begin{equation} \label{eq:decode_latency}
    \begin{split}
        T_{d} &= 
        l_{d}  
        (
        \frac{N(2 d_{h}d_{\textit{kv}} + d_{h}^2 + 3 d_{h}d_{\textit{ffn}}) + dd_{\textit{lm\_head}}}
        {\textit{WP}_{int4}} \\
        &+ \max(
        \frac{Nd_{h}^2}{\textit{WP}_{int4}},
        \frac{Nd_{h} (l_{p} + 0.5 l_{d})}
        {\textit{WP}_{\textit{mha}}}
        )
        )
    \end{split}
\end{equation}
\begin{equation}
BW_{d} = 
F(B_W^{\textit{int4}}\textit{WP}_{\textit{int4}} + 
2B_W^{\textit{int8}}\textit{WP}_{\textit{mha}})
\label{eq:decode_bandwidth}
\end{equation}


In practice, decode often uses larger \textit{WP} setting, creating oversized linear modules with high fan-out. We mitigate this by partitioning them into multiple identical submodules, improving floorplanning flexibility and achievable frequency during P\&R.

\begin{table}[t]
\centering
\scriptsize
\caption{Summary of module usage, code size, and development workload for each design component.}
\vspace{-0.2cm}
\renewcommand{\arraystretch}{1.1}
\setlength{\tabcolsep}{2.5pt}
\begin{tabular}{l|p{3.5cm}|c|c}
\hline
\textbf{Design} & \textbf{Main Module Usage} & \textbf{LoC} & \textbf{Workload to silicon} \\
\hline
Llama-3.2 1B & Linear, RoPE, MHA, KV\_cache, LayerNorm, Residual, Softmax, Swish, Gate, Sampling & $\sim600$ & $\sim$4 weeks $\times$ person \\
SpinQuant & Dyn. Asym. INT4 / Sta. Sym. INT8 Quant / Dequant, FHT & $\sim200$ & $\sim$1 weeks $\times$ person \\
HMT Plug-in & Linear, MHA, KV\_cache & $\sim200$ & $\sim$1 weeks $\times$ person \\
\hline
\end{tabular}
\label{tab:module_summary}
\vspace{-0.2cm}
\end{table}

\begin{table}[t]
  \centering
  \caption{Ablation Study on Llama-3.2-1B (lower PPL is better). }
  \label{tab:ablation}
  \footnotesize
  \setlength{\tabcolsep}{3pt}  
  \vspace{-0.2cm}
  \renewcommand{\arraystretch}{0.9}
  \resizebox{0.8\linewidth}{!}{
  \begin{tabular}{lcccccccccc}
    \toprule
    \textbf{Quant Config.} & \textbf{W} & \textbf{A} & \textbf{Attn} & \textbf{Vocab} &
    \textbf{Wiki-2 PPL} \\
    \midrule
    No\_Quant           & BF16 & BF16 & BF16        & BF16 & 8.94   \\
    Q0 (SpinQuant)      & INT4 & INT4 & BF16-INT4   & BF16 & 13.30  \\
    Q1                  & INT4 & INT4 & Dyn. INT8    & BF16 & 12.07  \\
    Q2                  & INT4 & INT4 & Sta. INT8    & BF16 & 12.28   \\
    Q3 (Final)          & INT4 & INT4 & Sta. INT8    & INT4 & 12.68 \\
    \bottomrule
  \end{tabular}}
\begin{minipage}{1.0\linewidth}
\scriptsize
\vspace{3pt}
\hspace{15pt}W/A/Attn/Vocab: weight/activation/attention/lm\_head precision.
\end{minipage}
\vspace{-0.8cm}
\end{table}

\begin{table*}[t]
  \centering
  \caption{Parameter configuration of the Llama~3.2-1B model, accelerator, and HMT plug-in, along with the on-board frequency, hardware resource utilization, and latency on the U280 FPGA and synthesis results (estimation marked with~*) on the V80 FPGA.}
  \label{tab:freq-util}
  \vspace{-0.2cm}
  \renewcommand{\arraystretch}{1.1}
  \setlength{\tabcolsep}{3.8pt}
  \resizebox{\textwidth}{!}{
  \begin{tabular}{l l c c c c c c c c}
    \hline
    \textbf{Model/Device} & \textbf{Design} & \textbf{Parameters} & \textbf{Freq.} & \textbf{CLB} & \textbf{DSP} & \textbf{LUT} & \textbf{FF} & \textbf{BRAM / URAM} & \textbf{Latency} \\
    \hline
    Llama 3.2-1B & -- & $L{=}16,\, d{=}2048,\, d_{kv}{=}512,\, d_{ffn}{=}8192,\, d_{lm\_head}{=}128256$ & -- & -- & -- & -- & -- & -- / -- & --\\
    \hline
    \multirow{3}{*}{U280} 
      & Prefill Arch. & $\textit{TP}{=}8,\, \textit{WP}_{\textit{kqvo}}{=}24,\, \textit{WP}_{\textit{mha}}{=}16,\, \textit{WP}_{\textit{ffn}}{=}96$ & 304\,MHz & 66\% & 29\% & 39\% & 24\% & 35\% / 11\% & 1.65\,s/1k tokens\\
      & Decode Arch.  & $\textit{BP}{=}16,\, \textit{WP}_{\textit{int4}}{=}1024,\, \textit{WP}_{\textit{mha}}{=}256$ & 292\,MHz & 76\% & 18\% & 44\% & 28\% & 41\% / 15\% & 6.94\,s/1k tokens\\
      & HMT Plug-in & $N{=}64,\, \textit{BP}{=}4,\, \textit{WP}_{\textit{mem\_attn}}{=}4$ & 290\,MHz & 7.5\% & 1.5\% & 5.3\% & 1.9\% & 4.3\% / 3.8\% & 8.44\,ms/segment\\
    \hline
    \multirow{2}{*}{V80}  
      & Prefill Arch. & $\textit{TP}{=}16,\, \textit{WP}_{\textit{kqvo}}{=}32,\, \textit{WP}_{\textit{mha}}{=}32,\, \textit{WP}_{\textit{ffn}}{=}128$ & 300\,MHz* & 58\% & 26\% & 37\% & 20\% & 22\% / 9\%  & 0.61\,s/1k tokens*\\
      & Decode Arch.  & $\textit{BP}{=}64,\, \textit{WP}_{\textit{int4}}{=}4096,\, \textit{WP}_{\textit{mha}}{=}1024$ & 300\,MHz* & 75\% & 25\% & 42\% & 22\% & 36\% / 20\% & 1.68\,s/1k tokens*\\
      & HMT Plug-in & $N{=}64,\, \textit{BP}{=}4,\, \textit{WP}_{\textit{mem\_attn}}{=}8$ & 300\,MHz* & 3.8\% & 0.7\% & 3.3\% & 0.9\% & 2.4\% / 1.9\% & 6.50\,ms/segment*\\
    \hline
  \end{tabular}}
\vspace{-0.5cm}
\end{table*}

\section{Case Study 2: HMT Integration}

While quantization reduces compute and memory traffic, long-context prompting still incurs quadratic attention cost in both computation and KV access, leading to high latency and energy. To address this, we integrate a Hierarchical Memory Transformer (HMT) plug-in into our accelerator system. HMT recurrently compresses context segments into compact memory embeddings and retrieves relevant information via a memory-attention pathway, reducing long-context processing complexity from quadratic to linear in sequence length. Moreover, this is an important study to showcase FlexLLM's flexibility and efficiency for implementing new LLM mechanisms.

FlexLLM’s composable design makes this integration lightweight: The HMT plug-in is built largely by reusing existing linear and attention modules to implement its segment encoder, memory generation, and history retrieval pipeline, requiring minimal additional hardware and design effort.

Fig.~\ref{fig:Implementation}(c) shows the end-to-end workflow. A long prompt is split into into multiple short segments. For each segment, the segment processor first forms a summary prompt by concatenating the first half of current segment ($Seg_n^T$) with a topic token embedding $T_n$, and sends it to the LLM accelerator to produce a topic summary vector $S_n$. The HMT plug-in then performs cross-attention between $S_n$ and the most recent $N$ memory embeddings $\{Mem_{n-N},\dots,Mem_{n-1}\}$ in a memory queue to generate a retrieved prompt embedding $P_n$.

Next, the HMT plug-in constructs an augmented prompt by concatenating the full segment, $P_n$, and a short-term context slice from the previous segment ($Seg_{n-1}^P$). This augmented input is processed by the LLM accelerator to produce the new long-term memory embedding $Mem_n$, which is appended to the memory queue for future retrieval.

The experimental results in \ref{results for HMT} show that the HMT plug-in built on FlexLLM enables scalable context window with minimal hardware overhead and negligible added latency over the backbone model implementation, substantially extending the applicability of LLM accelerators to long-context workloads.

\section{Evaluation}

\subsection{Experiment Setup}

\textbf{Hardware Platforms:}  
We select AMD Alveo U280~\cite{amd_u280_ds} at TSMC 16nm and Versal V80~\cite{amd_versal_v80_ds} at 7nm as our target accelerator platforms, and NVIDIA A100 at 7nm as GPU representative.
Table~\ref{tab:hardware_comparison} summarizes key hardware specifications.

\begin{figure}[t]
    \centering
    \includegraphics[width=\linewidth]{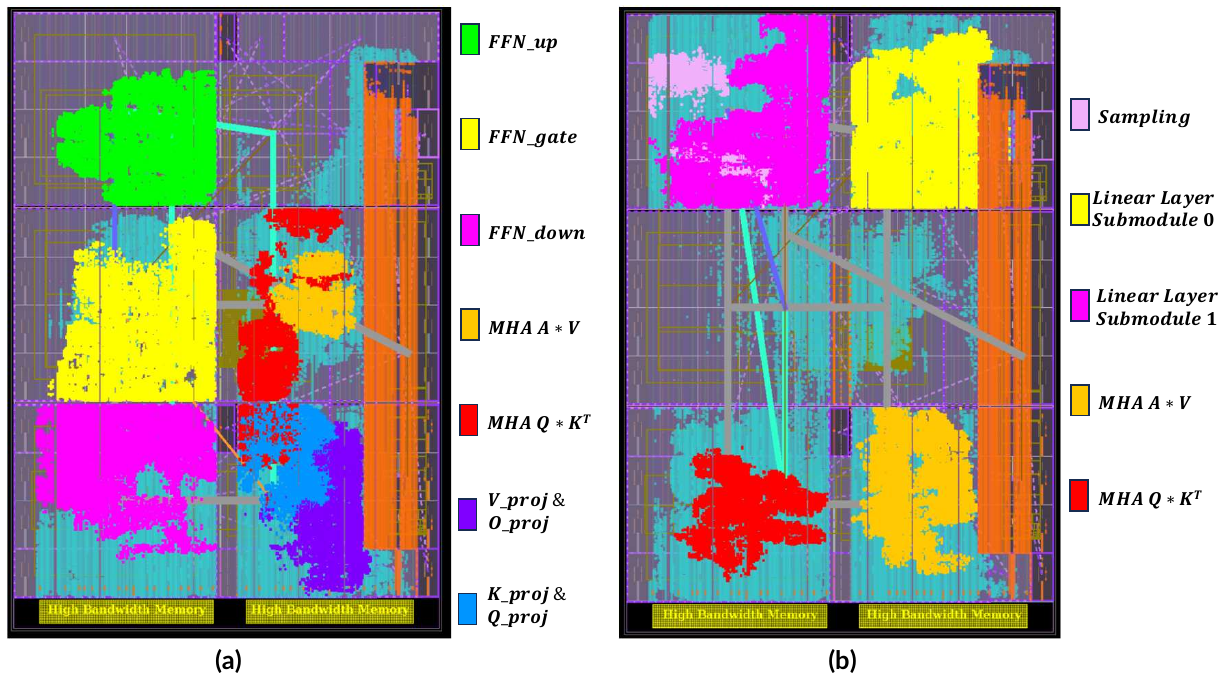}
    \vspace{-0.8cm}
    \caption{The implementation layout of (a) prefill and (b) decode architectures for quantized Llama-3.2 1B on U280.}
    \label{fig:Layout}
    \vspace{-0.6cm}
\end{figure}

\textbf{Accelerator Implementations:}  
 Using FlexLLM, we implement stage-customized accelerators for the Llama-3.2 1B model quantized with SpinQuant on both U280 and V80. Fig.~\ref{fig:Layout} shows the U280 layout. We also integrate the HMT plug-in for long-context processing. Table~\ref{tab:freq-util} summarizes model parameters, architectural configuration, achieved frequency, and resource utilization on both platforms. On U280, the design runs on-board with rapid reconfiguration ($\sim$0.3\,s) and performance measured directly on hardware. For V80 implementation, we perform RTL synthesis and P\&R using Vitis/Vivado~2024.2 and AVED toolchain, and project on-board performance by scaling from the U280 implementation. We use Allo~\cite{chen2024allo} with W4A8KV8 SmoothQuant as the SOTA accelerator baseline.

\textbf{GPU Baselines:}
We use the HuggingFace BF16 Llama-3.2 1B model as the GPU baseline. Since vLLM~\cite{kwon2023vllm} does not support SpinQuant, we instead quantize the model to INT4 using GPTQ~\cite{frantar2022gptq} with Marlin~\cite{frantar2025marlin} and evaluate it with vLLM on an A100 GPU. This setup provides a fair GPU reference for our FPGA implementations.

\textbf{Evaluation Metrics:}  
We evaluate FPGA-based accelerators and GPU baselines using three key metrics: (1) end-to-end latency, (2) decode throughput, and (3) energy efficiency. End-to-end latency measures the total inference time (prefill + decode), while decode throughput represents the number of tokens generated per second during the decoding. We record the average power from on-board sampling for the A100 and U280, and synthesis report for the V80,  and compute the tokens-per-joule (token/J) metric to evaluate energy efficiency.

\subsection{Experiment Results}

\begin{figure*}[t]
    \centering
    \includegraphics[width=1.0\linewidth]{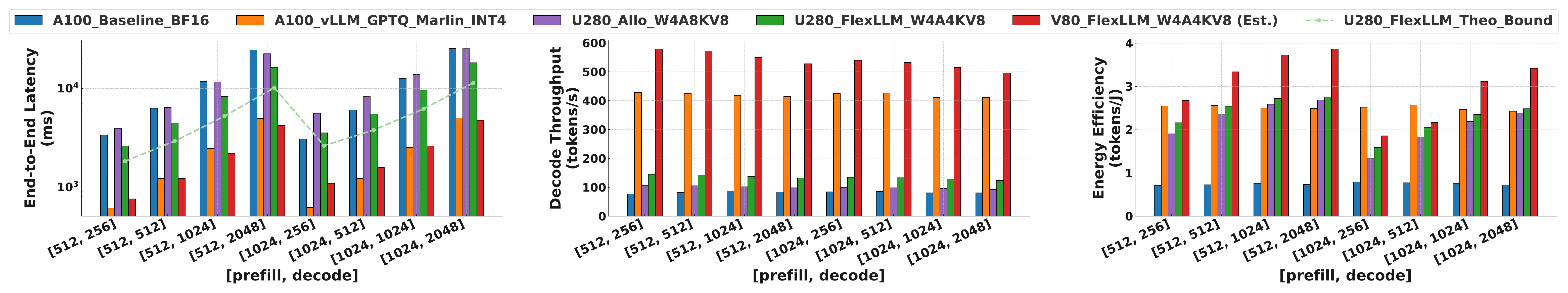}
    \vspace{-0.8cm}
    \caption{Performance and energy efficiency comparison of FlexLLM-based accelerators and A100 GPU under standard inference settings.}
    \label{fig:Perf_Comparison}
    \vspace{-0.2cm}
\end{figure*}

\begin{figure*}[t]
    \centering
    \includegraphics[width=1.0\linewidth]{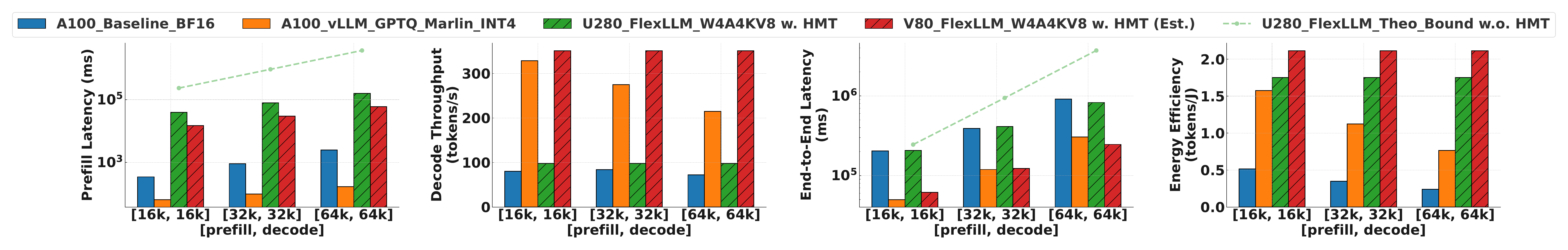}
    \vspace{-0.9cm}
    \caption{Performance and energy efficiency comparison in long-context scenarios with and without the HMT plug-in.}
    \label{fig:Perf_HMT}
    \vspace{-0.5cm}
\end{figure*}

\subsubsection{Case Study 1: Typical Sequence Lengths}  

We evaluate the proposed accelerators and the GPU baseline across a range of prefill and decode sequence lengths (Fig.~\ref{fig:Perf_Comparison}). As expected, the GPU demonstrates a clear advantage in scenarios with long prefill and short decode lengths (e.g., [1024, 256]), where its massive parallel compute capability yields ultra-fast prompt ingestion. However, as decode length grows, the performance bottleneck of the GPU shifts to memory bandwidth, while our FPGA accelerators sustain higher efficiency via stage-customized designs and flexible parallelism.

On average, the FlexLLM-based accelerator on U280 achieves \textbf{1.29$\times$} end-to-end speedup and \textbf{1.64$\times$} higher decode throughput over the A100 BF16 baseline, and also surpasses Allo with \textbf{1.46$\times$} and \textbf{1.35$\times$}, respectively. The V80 design further delivers up to \textbf{4.71$\times$} end-to-end speedup and \textbf{6.55$\times$} higher decode throughput. For longer decode workloads (e.g., [512, 2048] and [1024, 2048]), V80 even outperforms A100 running INT4 GPTQ-Marlin under vLLM. Although A100 has higher peak bandwidth, decode performance is bounded by effective bandwidth utilization: A100+vLLM achieves only 13.06\% average bandwidth utilization, whereas our V80 design reaches \textbf{38.61\%} on average and up to \textbf{74.93\%} peak. These results highlight the benefit of FlexLLM-enabled stage-customized architectures for memory-bound decode.

In energy efficiency, FlexLLM-based accelerators outperform the A100 BF16 baseline, achieving \textbf{3.14$\times$} (U280) and \textbf{4.13$\times$} (V80) higher tokens-per-joule on average. With long decode lengths (1024/2048), even the 16\,nm U280 surpasses the 7\,nm A100 running INT4 GPTQ-Marlin with vLLM, demonstrating strong utilization and power efficiency. Compared to Allo, our accelerator further achieves \textbf{1.10$\times$} higher energy efficiency. These results confirm that FlexLLM enables domain-specific accelerators that outperform general-purpose GPUs in both performance and energy efficiency, even across process nodes.

\subsubsection{Case Study 2: Long Context with HMT Plug-in}
\label{results for HMT}

In long-context inference, accelerators without HMT are prone to memory exhaustion due to limited FPGA HBM capacity. Even if off-chip memory suffices for KV caching, the theoretical prefill latency on U280 can exceed one hour at 64K tokens, making deployment impractical. This motivates sequence compression tailored for LLM accelerators, such as HMT.

As shown in Table~\ref{tab:freq-util}, the HMT plug-in on U280 consumes under \textbf{7.5\%} of total resources and adds only 8.44\,ms per segment (\textbf{\textless0.6\%} of end-to-end latency). FlexLLM’s templated modularity further enables tuning the plug-in’s resource--latency trade-off to match different LLM accelerators.

Figure~\ref{fig:Perf_HMT} reports performance and energy efficiency with HMT. The A100 without HMT remains faster than FPGA-based accelerators in prefill due to higher peak throughput; however, compared to the theoretical latency bound without HMT, HMT reduces prefill latency by up to \textbf{23.23$\times$}, bringing the time-to-first-token into a practical range for long prompts. For end-to-end performance, HMT-enhanced accelerators regain advantage over the A100, driven by higher decode throughput and HMT’s compression that converts the decode scaling from quadratic to linear in sequence length. Compared with the A100 BF16 baseline, the U280 achieves up to \textbf{1.36$\times$} higher decode throughput and \textbf{1.10$\times$} lower end-to-end latency, while the V80 reaches up to \textbf{4.86$\times$} and \textbf{3.70$\times$}, respectively.

HMT also stabilizes energy efficiency as context grows: tokens/J drops sharply on A100 for long prompts, whereas the HMT-enhanced U280 and V80 sustain high efficiency, delivering up to \textbf{5.21$\times$} and \textbf{6.27$\times$} higher tokens/J than the A100 BF16 baseline, and \textbf{1.65$\times$} and \textbf{1.99$\times$} over GPTQ-Marlin+vLLM. These results highlight not only the critical role of the HMT plug-in in enabling scalable long-context inference, but also the flexibility and extensibility of our FlexLLM framework for efficient LLM accelerator design.

\section{Conclusion and Outlook}
We present \textbf{FlexLLM}, a composable HLS framework for rapidly building and optimizing domain-specific LLM accelerators. FlexLLM enables stage-customized hybrid architectures for prefill and decode, and provides a comprehensive quantization stack for accurate low-bit inference, bridging model-level innovation and hardware-level implementation. Experimental results on the Llama-3.2 1B model demonstrate that FlexLLM enables high-performance and energy-efficient LLM inference designs, achieving performance comparable to or surpassing the NVIDIA A100 GPU running with vLLM. These results underscore FlexLLM’s capability to translate algorithmic innovations into efficient hardware realizations with minimal manual effort. Looking ahead, FlexLLM provides a foundation for domain-specific LLM acceleration on future platforms and for extending composable HLS methodologies to next-generation Transformer models.






\bibliographystyle{IEEEtran}

\bibliography{example_paper}
%



\end{document}